\documentclass[aip,
 jcp,%
 amsmath,amssymb,
 reprint,%
]{revtex4-1}
\usepackage{graphicx}
\usepackage{dcolumn}
\usepackage{bm}
\usepackage{amssymb}
\usepackage{array}
\usepackage{longtable}
\usepackage{color}

\begin{document}

\title{Optimal scale-free network with a minimum scaling of transport efficiency for random walks with a perfect trap}

\author{Yihang Yang}

\author{Zhongzhi Zhang}
\email{zhangzz@fudan.edu.cn}
\homepage{http://homepage.fudan.edu.cn/~zhangzz/}

\affiliation {School of Computer Science, Fudan University,
Shanghai 200433, China}

\affiliation {Shanghai Key Lab of Intelligent Information
Processing, Fudan University, Shanghai 200433, China}

\begin{abstract}
Average trapping time (ATT) is central in the trapping problem since it is a key indicator characterizing the efficiency of the problem. Previous research has provided the scaling of a lower bound of the ATT for random walks in general networks with a deep trap. However, it is still not well understood in which networks this minimal scaling can be reached. Particularly, explicit quantitative results for ATT in such networks, even in a specific network, are lacking, in spite that such networks shed light on the design for optimal networks with the highest trapping efficiency. In this paper, we study the trapping problem taking place on a hierarchical scale-free network with a perfect trap. We focus on four representative cases with the immobile trap located at the root, a peripheral node, a neighbor of the root with a single connectivity, and a farthest node from the root, respectively. For all the four cases, we obtain the closed-form formulas for the ATT, as well as its leading scalings. We show that for all the four cases of trapping problems, the dominating scalings of ATT can reach the predicted minimum scalings. This work deepens the understanding of behavior of trapping in scale-free networks, and is helpful for designing networks with the most efficient transport process.
\end{abstract}



\date{\today}
\maketitle

\section{Introduction}

As a paradigmatic random walk, trapping problem has received increasing interest within the scientific community, which was first introduced in the pioneering work by Montroll more than forty years ago~\cite{Mo69}. It is a kind of random walk in graphs in the presence of a single deep trap positioned at a given location, absorbing all particles (walkers) that visit it. Trapping process is relevant in a large variety of other dynamical processes occurring in a number of different complex systems. Frequently cited examples include light harvesting in antenna systems~\cite{BaKlKo97,BaKl98,Ag11}, energy or exciton transport in polymer systems~\cite{BlZu81,SoMaBl97,HeMaKn04,BaKl98JCP}, page search or access in the World Wide Web~\cite{HwLeKa12,HwLeKa12E}, to name a few. Because of its practical relevance, it is of significant importance to address trapping problem in diverse complex systems.

A fundamental quantity pertaining to trapping problem is the trapping time (TT), often referred to as the mean first-passage time (MFPT)~\cite{Re01,MeKl04,NoRi04,BuCa05,CoBeTeVoKl07}. The TT for a node $i$ is the expected time taken by a particle starting off from $i$ to first visit the trap. The average trapping time (ATT)
is defined as the average of TT over all possible starting nodes in the system, which  can be used as an indicator of the efficiency of trapping. In the past few years, trapping problem in different kinds of graphs has been intensively studied, including the square and cubic lattices~\cite{GLKo05,GLLiYoEvKo06}, dendrimers~\cite{BeHoKo03,BeKo06,WuLiZhCh12}, the Sierpinski gasket~\cite{KaBa02PRE,MeAgBeVo12} and Sierpinski tower~\cite{KaBa02IJBC,BeTuKo10}, the $T-$fractal~\cite{KaRe89,Ag08,HaRo08} and its extension~\cite{LiWuZh10,ZhWuCh11}, as well as scale-free graphs~\cite{KiCaHaAr08,ZhQiZhXiGu09,AgBu09,TeBeVo09,ZhLiGoZhGuLi09,ZhYaLi12}.
These studies showed how the ATT scales with the size of systems with various topological properties, unveiling the nontrivial effects of graph structure on the behavior of the ATT.

More recently, it has been proven~\cite{LiJuZh12} that for trapping problem in any graph with a trap fixed at an arbitrary node, the possible minimal scaling for the ATT is proportional to the graph size and the inverse degree of the trap node, which is universal and provides a maximal scaling for the lower bound of ATT for trapping in an arbitrary network with an immobile trap. Thus, a network is called optimal if this minimal scaling can be achieved for any node as a trap. However, since the MFPT from one node to another depends on the source-target distance~\cite{TeBeVo11}, for trapping in a graph, the fact that the minimal scaling can be reached when the trap is fixed on one trap does not necessarily lead to the conclusion that the minimal scaling can be obtained when the trap is located at another node. For example, for trapping in Cayley trees~\cite{WuLiZhCh12} as a model of polymer networks~\cite{GuBl05,CaCh97,ChCa99,BiKaBl01,MuBiBl06}, the minimal scaling can be reached when the trap is at the core node, but it cannot be achieved if the trap is at a boundary node. Thus, it is interesting to design or find optimal graphs, where the minimal scaling for ATT can be reached for any trap node. In particular, it is useful to derive closed-form solutions to ATT for separate trap nodes having minimal scaling since they are instrumental to quantitative understanding of theoretical models and to approximate or numerical solutions.

In this paper, we study the trapping problem in a hierarchical scale-free network~\cite{BaRaVi01}. We study four cases of trapping problems with the immobile trap being fixed at four representative nodes, i.e., the root node, a peripheral node, a neighboring node of the root with a single degree, and a farthest node from the root, respectively. For all these four cases, we derive analytically the explicit expressions for the ATT, based on which we further obtain their leading scalings for large network sizes. We show that the four trapping processes are all very efficient since their ATT grows linearly with the network size for the worst case. Moreover, we show that for the four representative cases of trapping problems, the possible minimum scalings for ATT can be achieved. In this sense, the network being studied is optimal for trapping processes, which is helpful for designing networks with the highest trapping efficiency.

\section{Construction and properties of the hierarchical scale-free network}

\begin{figure}
\begin{center}
\includegraphics[width=0.85\linewidth,trim=00 00 0 0]{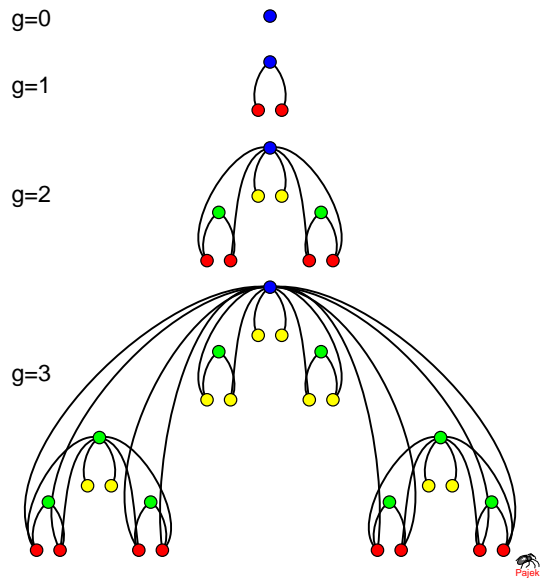}
\end{center}
\caption[kurzform]{\label{network} (Color online) Iterative
construction process of the hierarchical scale-free network. The red nodes are peripheral nodes.}
\end{figure}

We first introduce the model of the hierarchical scale-free network~\cite{BaRaVi01}, which is constructed in an iterative manner. Let $G_{g}$ $(g\geq0)$ denote the network after $g$ iterations. Initially $(g=0)$, the network $G_{0}$ contains one node without any edge, which is called main hub (or root) node. At iteration $g=1$, to generate $G_{1}$ we introduce two more nodes and link them to the original node in $G_{0}$. The two new nodes are named peripheral nodes of $G_{1}$, and the root node of $G_{0}$ is also the root of $G_{1}$. For $g\geq1$, $G_{g}$ is obtained from $G_{g-1}$ by adding two new copies, denoted by $G_{g-1}^{(1)}$ and $G_{g-1}^{(2)}$, of $G_{g-1}$ to the primal $G_{g-1}$, with all peripheral nodes of the two replicas being linked to the main hub of the original $G_{g-1}$ unit. The hub of the original $G_{g-1}$ and the peripheral nodes of the two duplicates of $G_{g-1}$ form the main hub node and peripheral nodes of $G_{g}$, respectively. Repeating indefinitely the two steps of replication
and connection leads to the hierarchical scale-free network.
Figure~\ref{network} illustrates schematically the network $G_3$.

It is easy to obtain that the number of nodes in $G_{g}$, denoted by $N_{g}$, is $N_{g}=3^{g}$. According to the node degree, all these nodes can be separated into four different sets by applying the method in~\cite{No03}: the peripheral node set $\mathbb P$, the locally peripheral node set ${\mathbb P}_{c}$ $(1\leq c < g)$, the root node set $\mathbb H$ comprising only the main hub of $G_{g}$, and the local hub set ${\mathbb H}_{c}$ $(1\leq c < g)$. For detailed explanation, we refer to reference~\cite{No03}.  All nodes in a set have the same degree, and the degree for a node in sets $\mathbb P$, ${\mathbb P}_c$, $\mathbb H$, and ${\mathbb H}_c$ is,
\begin{equation}\label{II5}
K_{p}(g)=g,
\end{equation}
\begin{equation}\label{II6}
K_{p,c}(g)=c,
\end{equation}
\begin{equation}\label{II7}
K_{h}(g)=\sum_{i=1}^{g}2^i=2(2^g-1),
\end{equation}
and
\begin{equation}\label{II8}
K_{h,c}(g)=\sum_{i=1}^{c}2^i=2(2^c-1)\,,
\end{equation}
respectively. In addition, the number of nodes in each of these four sets is, respectively,
\begin{equation}\label{II1}
|{\mathbb P}|=2^{g},
\end{equation}
\begin{equation}\label{II2}
|{\mathbb P}_{c}|=2^{c}3^{g-c-1},
\end{equation}
\begin{equation}\label{II3}
|{\mathbb H}|=1,
\end{equation}
and
\begin{equation}\label{II4}
|{\mathbb H}_{c}|=2\times3^{g-c-1}\,.
\end{equation}
Based on the above quantities, it is easy to derive that the average degree of all nodes is
\begin{equation}\label{II9}
\bar{K}(g)=4\left[1-\left(\frac{2}{3}\right)^g\right],
\end{equation}
which is approximately equal to a constant 4 when $g$ is large enough, implying that the network is sparse.

In $G_g$, the maximal distance from the root to any other node equals to $g$. Let ${\mathbb F}_g$ denote the set of those nodes in $G_g$ at a distance $g$ from the main hub. We call such nodes the farthest nodes of $G_g$. As shown above, $G_g$ is composed of a primal $G_{g-1}$ and two replicas of $G_{g-1}$, i.e., $G_{g-1}^{(1)}$ and $G_{g-1}^{(2)}$. The farthest nodes of $G_g$ must be in the two subgraphs $G_{g-1}^{(1)}$ and $G_{g-1}^{(2)}$. Specifically, ${\mathbb F}_g$ contains the  farthest nodes of the primal central subgraph, i.e., $G_{g-2}$, forming $G_{g-1}^{(1)}$ and $G_{g-1}^{(2)}$. Then, the number of nodes in ${\mathbb F}_g$, denoted by $|{\mathbb F}_{g}|$, satisfies the following recursive relation:
\begin{equation}\label{L1}
|{\mathbb F}_{g}|=2|{\mathbb F}_{g-2}|.
\end{equation}
Considering the initial conditions $|{\mathbb F}_1|=2$ and $|{\mathbb F}_2|=2$, Eq.~(\ref{L1}) is solved to yield
\begin{equation}\label{L2}
|{\mathbb F}_{g}|=
\begin{cases}
2^{(g+1)/2}, &g \quad {\rm is \quad odd}, \\
2^{g/2}, &g \quad {\rm is \quad even}.
\end{cases}
\end{equation}

The network being studied presents some typical properties as observed in many real-life
systems~\cite{BaRaVi01,IgYa05}. It is scale-free~\cite{BaAl99} with its degree
distribution $P(k)$ following a power law form $P(k) \sim k^{-\gamma}$, where the exponent $\gamma=1+\frac{\ln 3}{\ln 2}$. It displays the small-world behavior~\cite{WaSt98} with its average distance increasing logarithmically with its size~\cite{ZhLiGaZhGu09}. Moreover, the network has an
obvious hierarchical structure that has also been observed in real-world networks, e.g., metabolic networks~\cite{RaSoMoOlBa02}. It is the precursor,
probably the first model for hierarchical scale-free networks. It is thus
of great interest to study how the unique structure affects dynamical processes taking place on the network.

\section{Trapping in the hierarchical scale-free network}

After introducing the model and properties of the hierarchical scale-free network, in this
section we study analytically a particular random walk---the trapping problem---occurring on $G_{g}$, where a deep trap is positioned at a certain node.

We first consider the simple random-walk model on a general network with $N$ nodes. At each time step, the walker jumps from its current position to any of its neighboring nodes with an identical probability. Assume that node $j$ is trap node. Let $T_{ij}$ denote the trapping time for node $i$, i.e., MFPT from node $i$ to node $j$, which is the expected
time for a walker staring off from node $i$ to first
visit the trap $j$. The highly desirable
quantity related to the trapping problem is the ATT, $T_j$, which is the average of $T_{ij}$ over all the $N$ source nodes distributed uniformly over the whole network. By
definition, $T_j$ is given by
\begin{equation}\label{ATT}
T_j =\frac{1}{N}\sum_{i=1}^{N} T_{ij}\,.
\end{equation}

In the sequel, in order to explore the impacts of the trap position and its degree, we will study analytically four cases of trapping problem performed on the hierarchical scale-free network $G_{g}$, with the perfect trap located at the root node, a peripheral node, a neighboring node of root node having a single degree, and a farthest node, respectively. For these four representative trapping problems, we will explicitly determine the ATT and show how their dominating behaviors scale with the network size.

\subsection{Trapping with the trap being positioned at the root node}

We first address the case when the trap is fixed at the root node of $G_g$. To this end, we define a multiple trap problem with all the peripheral nodes being the traps. Let $T_h(g)$ denote the ATT to the root node; and let the $T_p(g)$ denote the ATT when all peripheral nodes are occupied  by traps.

In order to evaluate $T_h(g)$ and $T_p(g)$, we introduce two intermediate variables $T_{p,h}(g)$ and $T_{h,p}(g)$. The former represents the MFPT from an arbitrary peripheral node to the root node of $G_g$, while the latter $T_{h,p}(g)$ stands for the MFPT from the root node to any of the $2^g$ arbitrary peripheral nodes in $G_g$.
In Appendix~\ref{AppA}, we derive analytically the two intermediate variables $T_{p,h}(g)$ and $T_{h,p}(g)$, which read
\begin{equation}\label{II10}
T_{p,h}(g)=\frac{8}{3}\left(\frac{3}{2}\right)^{g}-3
\end{equation}
and
\begin{equation}\label{II11}
T_{h,p}(g)=\frac{4}{3}\left(\frac{3}{2}\right)^{g}-1\,,
\end{equation}
respectively. It should be mentioned that Eqs.~(\ref{II10}) and~(\ref{II11}) were first found in~\cite{AgBu09} by using the approach of generating functions.
The obtained results for $T_{p,h}(g)$ and $T_{h,p}(g)$ are very useful quantities for the following calculations.

Having obtained the intermediate quantities, we now determine the quantities $T_h(g)$ and $T_p(g)$. From the network structure, the following recursive relations can be established:
\begin{equation}\label{II12}
T_{h}(g)=\frac{1}{3}T_{h}(g-1)+\frac{2}{3}\left[T_{p}(g-1)+T_{p,h}(g)\right]
\end{equation}
and
\begin{equation}\label{II13}
T_{p}(g)=\frac{1}{3}\left[T_{h}(g-1)+T_{h,p}(g)\right]+\frac{2}{3}T_{p}(g-1).
\end{equation}
After some algebra, Eqs.~(\ref{II12}) and~(\ref{II13}) can be recast as
\begin{equation}\label{II14}
3T_{h}(g)-T_{h}(g-1)=2\left[T_{p}(g-1)+T_{p,h}(g)\right]
\end{equation}
and
\begin{equation}\label{II15}
3T_{p}(g)-2T_{p}(g-1)-T_{h,p}(g)=T_{h}(g-1).
\end{equation}
From Eq.~(\ref{II15}), we can further have
\begin{equation}\label{II16}
3T_{p}(g+1)-2T_{p}(g)-T_{h,p}(g+1)=T_{h}(g),
\end{equation}
which, together with Eq.~(\ref{II15}), gives
\begin{eqnarray}\label{II17}
&\quad&3\left[3T_{p}(g+1)-2T_{p}(g)-T_{h,p}(g+1)\right]\nonumber\\
&\quad&-\left[3T_{p}(g)-2T_{p}(g-1)-T_{h,p}(g)\right]\nonumber\\
&=&3T_{h}(g)-T_{h}(g-1)=2\left[T_{p}(g-1)+T_{p,h}(g)\right],
\end{eqnarray}
namely,
\begin{equation}\label{II18}
T_{p}(g+1)=T_{p}(g)+\frac{1}{9}\left[3T_{h,p}(g+1)-T_{h,p}(g)+2T_{p,h}(g)\right].
\end{equation}
Inserting Eqs.~(\ref{II10}) and~(\ref{II11}) into Eq.~(\ref{II18}) and considering the initial condition $T_{p}(1)=1/3$, we can solve Eq.~(\ref{II18}) to obtain the explicit expression for the ATT when all peripheral nodes are occupied by traps:
\begin{equation}\label{II19}
T_{p}(g)=\frac{20}{9}\left(\frac{3}{2}\right)^{g}-\frac{8}{9}g-\frac{19}{9}\,,
\end{equation}
which is consistent with the previous result in~\cite{AgBuMa10}. Substituting Eq.~(\ref{II19}) into Eq.~(\ref{II14}) and considering the initial value $T_{h}(1)=2/3$, we can solve Eq.~(\ref{II14}) to obtain the exact formula for the ATT to the root node:
\begin{equation}\label{II20}
T_{h}(g)=\frac{32}{9}\left(\frac{3}{2}\right)^{g}-\frac{8}{9}g-\frac{34}{9}.
\end{equation}
Notice that Eq.~(\ref{II20}) agrees with the result derived in~\cite{AgBu09} by using the approach of generating functions.

We proceed to express $T_{p}(g)$ and $T_{h}(g)$ in terms of network size $N_{g}$, with an aim to obtain the dependence of the two quantities on $N_{g}$. Recalling $N_{g}=3^g$, we have $g=\ln N_{g}/ \ln3$ and $2^g= (N_{g})^{\ln2/ \ln3}$, which allow to represent Eqs.~(\ref{II19}) and~(\ref{II20}) as a function of $N_{g}$ as
\begin{equation}\label{II21}
T_{p}(g)=\frac{20}{9}(N_{g})^{1-\ln2/ \ln3}-\frac{8}{9}\frac{\ln N_{g}}{\ln3}-\frac{19}{9}
\end{equation}
and
\begin{equation}\label{II22}
T_{h}(g)=\frac{32}{9}(N_{g})^{1-\ln2/ \ln3}-\frac{8}{9}\frac{\ln N_{g}}{\ln3}-\frac{34}{9}.
\end{equation}
When the system is very large, i.e., $N_{g}\to \infty$,
\begin{equation}\label{II23}
T_{h}(g)\sim (N_{g})^{1-\ln2/ \ln3}
\end{equation}
and
\begin{equation}\label{II24}
T_{p}(g)\sim (N_{g})^{1-\ln2/ \ln3},
\end{equation}
both of which scale sublinearly with the network size. We note that Eqs.~(\ref{II23}) and~(\ref{II24}) have been perviously derived in~\cite{AgBuMa10} by using another technique, which is different from the one adopted here.

Since the degree of the root node is $K_{h}(g)=2(2^g-1)$, which is can be expressed in term of network size $N_{g}$ as $K_{h}(g) \sim (N_{g})^{\ln2/ \ln3}$ when $N_{g}$ is very large~\cite{IgYa05}, then $T_{h}(g)$ can be expressed as a function of $N_{g}$ and $K_{h}(g)$ as
\begin{equation}\label{II233}
T_{h}(g)\sim \frac{N_{g}}{K_{h}(g)}\,,
\end{equation}
which grows proportionally to the network size and inverse degree of the root node as the trap.

\subsection{Trapping with the trap being located at a peripheral node}

We now consider the second case that the trap is fixed at one of the peripheral nodes. As shown above, there are $|{\mathbb P}|=2^g$ peripheral nodes in $G_g$. We label sequentially these peripheral nodes by $1,2,\cdots,2^g$ from left to right. For convenience, we classify all the peripheral nodes into $g+1$ sets denoted by $\beta_{i}$ ($0 \leq i \leq g$): For $i=0$, $\beta_0=\{1\}$; while for $1 \leq i \leq g$, $\beta_{i}=\{x_i|2^{i-1}<x_i\leq2^i\}$. In addition, let ${\mathcal B}_i (0\leq i \leq g)$ be the union of the sets $\beta_{k}$ with $0 \leq k \leq i$, namely ${\mathcal B}_i=\bigcup_{k=0}^{i}\beta_k$. Since for this particular trapping problem, the selection of trap position has no effect on the ATT, with loss of generality, we choose the node belonging to $\beta_0$ as the trap and denote $T_{{\mathcal B}_0}(g)$ as the ATT.

Prior to deriving the ATT $T_{{\mathcal B}_0}(g)$, we define and determine some new quantities. Let $T_{\beta_{i+1},{\mathcal B}_i}(g)$ denote the MFPT for a walker starting from an arbitrary node in $\beta_{i+1}$ to an arbitrary node belonging to $\mathcal B_i$ in $H_{g}$.  Then, we have the following relation:
\begin{eqnarray}\label{III2}
T_{\beta_{i+1},{\mathcal B}_i}(g)&=&\frac{1}{g}\sum_{k=1}^{i}\left[1+T_{h,p}(k)+T_{\beta_{i+1},{\mathcal B}_i}(g)\right]+\nonumber\\
&\quad&\frac{1}{g}\sum_{k=i+1}^{g}\left[1+T_{h,p}(k)+\frac{1}{2}\sum_{l=i}^{k-1}T_{\beta_{l+1},{\mathcal B}_l}(g)\right].\nonumber\\
\end{eqnarray}
The two terms on the rhs of Eq.~(\ref{III2}) can be elaborated as follows: the first term explains the process that a walker starting from a node in $\beta_{i+1}$ takes one step to reach a local hub node that has no links to other peripheral nodes except those in ${\beta}_{i+1}$, and then jumps $T_{h,p}(k)+T_{\beta_{i+1},{\mathcal B}_i}(g)$ more steps to hit a target. The second term describes the process that the walker starting from $\beta_{i+1}$ takes one step to reach a local hub that has a link connected to peripheral nodes not in ${\beta}_{i+1}$, then takes $T_{h,p}(k)+\frac{1}{2}\sum_{l=i}^{k-1}T_{\beta_{l+1},{\mathcal B}_k}(g)$ steps to visit a destination.

From Eq.~(\ref{III2}), we can derive
\begin{small}
\begin{equation}\label{III3}
T_{\beta_{i+1},{\mathcal B}_i}(g)=\frac{2}{g-i}\left[g+\sum_{k=1}^{g}T_{h,p}(k)+\frac{1}{2}\sum_{k=i+1}^{g}\sum_{l=i+1}^{k-1}T_{\beta_{l+1},{\mathcal B}_l}(g)\right]
\end{equation}
\end{small}
and
\begin{small}
\begin{equation}\label{III4}
T_{\beta_{i},{\mathcal B}_{i-1}}(g)=\frac{2}{g-i+1}\left[g+\sum_{k=1}^{g}T_{h,p}(k)+\frac{1}{2}\sum_{k=i}^{g}\sum_{l=i}^{k-1}T_{\beta_{l+1},{\mathcal B}_l}(g)\right],
\end{equation}
\end{small}
which, together with Eq.~(\ref{II11}), leads to
\begin{equation}\label{III5}
T_{\beta_i,{\mathcal B}_{i-1}}(g)=\frac{2(g-i)}{g-i+1}T_{\beta_{i+1},{\mathcal B}_i}(g).
\end{equation}
Using the initial condition $T_{\beta_g,{\mathcal B}_{g-1}}(g)=8(3/2)^g-8$, we can solve Eq.~(\ref{III5}) to obtain
\begin{equation}\label{III6}
T_{\beta_{i+1},{\mathcal B}_i}(g)=\frac{4(3^g-2^g)}{(g-i)2^i}\,.
\end{equation}

The above obtained quantity $T_{\beta_{i+1},{\mathcal B}_i}(g)$ enables us to obtain the ATT $T_{{\mathcal B}_0}(g)$. By construction, we have the following relation
\begin{small}
\begin{eqnarray}\label{III10}
T_{{\mathcal B}_0}(g)&=& \frac{1}{3} \left[T_{h}(g-1)+T_{h,p}(g)+\frac{1}{2}\sum_{k=0}^{g-1}T_{\beta_{k+1},{\mathcal B}_k}(g)\right]+\nonumber\\
&\quad&\frac{1}{3}\bigg[T_{p}(g-1)+T_{\beta_g,{\mathcal B}_{g-1}}(g)+\frac{1}{2}\sum_{k=0}^{g-2}T_{\beta_{k+1},{\mathcal B}_k}(g)\bigg]+\nonumber\\
&\quad& \frac{1}{3^g}\bigg\{\sum_{i=2}^{g-1}3^{i-1}\bigg[T_{h}(i-1)+T_{h,p}(i)+\frac{1}{2}\sum_{k=0}^{i-1}T_{\beta_{k+1},{\mathcal B}_k}(g)\nonumber\\
&\quad&+T_{p}(i-1)+T_{\beta_i,{\mathcal B}_{i-1}}(g)+\frac{1}{2}\sum_{k=0}^{i-2}T_{\beta_{k+1},{\mathcal B}_k}(g)\bigg]\nonumber\\
&\quad&+\left[T_{h,p}(1)+\frac{1}{2}T_{\beta_1,{\mathcal B}_0}(g)+T_{\beta_1,{\mathcal B}_0}(g)\right] \bigg\}\,.
\end{eqnarray}
\end{small}
The three terms on the rhs of Eq.~(\ref{III10}) can be accounted for as follows. Note that $G_g$ consists of three copies of $H_{g-1}$: central $G_{g-1}$, $G_{g-1}^{(1)}$, and $G_{g-1}^{(2)}$. The first term on the rhs of Eq.~(\ref{III10}) describes the contribution to ATT made by nodes in the central $G_{g-1}$.  The second term presents the trapping time of nodes in the subgraphs $G_{g-1}^{(2)}$. The third term is a little complicated, which explains the trapping time of nodes in the subgraph $G_{g-1}^{(1)}$.

Inserting the above obtained related quantities into Eq.~(\ref{III10}), we can obtain the explicit expression for the ATT $T_{{\mathcal B}_0}(g)$ as
\begin{eqnarray}\label{III11}
T_{{\mathcal B}_0}(g)&=&\frac{1}{3^g} + \frac{6}{g}\left[1-\left(\frac{2}{3}\right)^g\right]+\sum_{i=1}^{g-1}\frac{1}{3^i}\Bigg[ \frac{70}{9} \left( \frac{3}{2} \right)^{g-i}\nonumber\\
&\quad&- \frac{16}{9}(g-i) - \frac{62}{9}+4\left[\left(\frac{3}{2}\right)^g-1\right]\sum_{j=i+1}^{g} \frac{2^j}{j}\nonumber\\
&\quad& + \frac{6\left(3^g-2^g\right)}{2^{g-i}i}\Bigg].
\end{eqnarray}
It is not difficult to find that the term with the highest exponent occurs when $i=1$ and $j=g$. Moreover, in the infinite network size limit, i.e., $N_g\to\infty$, we have
\begin{equation}\label{III12}
T_{{\mathcal B}_0}(g)\sim N_g/\log_3 N_g,
\end{equation}
which, together with Eqs.~(\ref{II24}), indicates that when the trap is fixed at a peripheral node the trapping process displays a lower efficiency than the case when all peripheral nodes are traps. Thus, the number of traps sensitively affects the behavior of trapping processes taking place on the hierarchical scale-free network. Based on $N_{g}=3^{g}$ and Eq.~(\ref{II5}), we have $K_{p}(g)=g=\log_3 N_g$, which together with Eqs.~(\ref{III12}) leads to
\begin{equation}\label{III122}
T_{{\mathcal B}_0}(g)\sim N_g/K_{p}(g)\,,
\end{equation}
implying that $T_{{\mathcal B}_0}(g)$ is proportional to network size and the inverse degree of the trap node.

\subsection{Trapping with the trap being placed at a root's neighbor with a single degree}

Here we address the third trapping problem with the trap being positioned at a neighbor of the root, which has only one edge. Note that among all neighbors of the root, there are only two neighbors having a single connectivity, denoted by $x$ and $y$, respectively. For convenience, let $x$ be the trap node. The quantity we are concerned with is the ATT to trap node $x$ for trapping in $G_{g}$, denoted by $T_{x}(g)$. In order to determine $T_{x}(g)$, we first determine the MFPT, $T_{h,x}(g)$, from the main hub to node $x$ in $G_{g}$
By construction, $T_{h,x}(g)$ can be written recursively as
\begin{eqnarray}\label{IV1}
T_{h,x}(g)&=&\frac{1}{2(2^g-1)}+\frac{1}{2(2^g-1)}[1+1+T_{h,x}(g)]+\nonumber\\
&\quad&\frac{1}{2(2^g-1)}\sum_{i=2}^{g}2^{i}[1+T_{p,h}(i)+T_{h,x}(g)].
\end{eqnarray}

The first term on the rhs of Eq.~(\ref{IV1}) describes the fact that the walker, starting from the main hub, requires only one time step to hit the trap node $x$. The second term explains the process by which the walker first jumps to node $y$ in one time step, then takes one step to return to the root, and continues to jump $T_{h,x}(g)$ more steps to reach the target. The last term accounts for the fact that the walker first makes a jump to a peripheral node or local peripheral node belonging to ${\mathbb P}$ or ${\mathbb P}_{i}$ ($2\leq i \leq g$), then takes $T_{p,h}(i)$ time steps to the hub, and proceeds to the node $x$, taking $T_{h,x}(g)$ more time steps.

Equation~(\ref{IV1}) can be simplified to
\begin{equation}\label{IV2}
T_{h,x}(g)=1+\sum_{i=1}^{g}2^i+\sum_{i=2}^{g}2^{i}T_{p,h}(i).
\end{equation}
Considering the initial condition $T_{h,x}(1)=3$ and Eq.~(\ref{II10}), we can solve Eq.~(\ref{IV2}) to produce the exact solution for $T_{h,x}(g)$, which reads
\begin{equation}\label{IV3}
T_{h,x}(g)=4(3^g-2^g)-1\,.
\end{equation}
Note that the expression in Eq.~(\ref{IV3}) is actually a special case of the more general formula derived in a different approach in~\cite{MeAgBeVo12}. Then, the quantity $T_{x}(g)$ can be accurately evaluated as
\begin{eqnarray}\label{IV4}
T_{x}(g)&=&T_{h}(g)-\frac{1}{N_g}+\frac{N_g-1}{N_g}T_{h,x}(g)\nonumber\\
&=&4(3^g-2^g)+\frac{32}{9}\left(\frac{3}{2}\right)^{g}+4\left(\frac{2}{3}\right)^{g}-\frac{8}{9}g-\frac{79}{9},\nonumber\\
\end{eqnarray}
which can be expressed in terms of network size $N_g$ as
\begin{eqnarray}\label{IV5}
T_{x}(g)&=&4\,N_g-4\,(N_{g})^{\ln2/ \ln3}+\frac{32}{9}(N_{g})^{1-\ln2/ \ln3}+\nonumber\\
&\quad&4(N_{g})^{\ln2/ \ln3-1}-\frac{8}{9}\frac{\ln N_{g}}{\ln3}-\frac{79}{9}.
\end{eqnarray}
Thus, when $N_g\to\infty$,
\begin{equation}\label{IV6}
T_{x}(g) \sim N_g,
\end{equation}
that is, the leading term of $T_{x}(g)$ grows linearly with the network size but inversely proportional to the trap's degree, which is exactly equal to 1 in this particular case.

\subsection{Trapping with the trap being fixed at a farthest node}

Now we focus on the case when the trap is located at a farthest node in $G_g$. Since in this case, the ATT does not depend on the location of the trap, without loss of generality, we choose the leftmost farthest node as the target that belongs to $G_g^{(1)}$, and we let $T_{f}(g)$ be the ATT for this special trapping problem. Below, we first concentrate on the MFPT from the main hub to the trap in $G_g$, denoted by $T_{h,f}(g)$, based on which we will determine $T_{f}(g)$. Furthermore we will show that both $T_{h,f}(g)$ and $T_{f}(g)$ have the same leading scaling.

\subsubsection{Related definitions and quantities}

In order to derive the expressions for $T_{h,f}(g)$ and $T_{f}(g)$, we introduce some more variables. For those nodes of $G_g$ that belong to $G_{g-1}^{(1)}$ or $G_{g-1}^{(2)}$, we can classify them in the following way. Let ${\mathcal H}_{g-i} (0\leq i \leq g-1)$ be the set of the local hub nodes that are directly connected to $g-i$ classes of local peripheral nodes in ${\mathbb P}_{c}$, and let ${\mathcal P}_{g-i}$ $(0\leq i\leq g-1)$ denote the set of the local peripheral nodes whose neighbors are $g-i$ different local hub nodes belonging to ${\mathbb H}_{c}$. In addition, we assume that ${\mathcal H}_g = {\mathbb H}$ and ${\mathcal P}_g={\mathbb P}$.

The specific structure of the network shows that for a particle starting from the main hub to one of the $|{\mathbb F}_{g}|$ farthest nodes, it must follow the path ${\mathcal H}_{g} \to {\mathcal P}_{g} \to {\mathcal H}_{g-1} \to {\mathcal P}_{g-2} \to {\mathcal H}_{g-3} \to \cdots \to {\mathcal P}_{g-(i-1)} \to {\mathcal H}_{g-i} \to {\mathcal P}_{g-(i+1)} \to {\mathcal H}_{g-(i+2)} \to \cdots \to {\mathcal H}_{1}~{\rm or}~{\mathcal P}_{1}$. For the particular case that the leftmost farthest node is chosen as the target, the path should be definitely as follows: each time the walker starting from a current local hub in ${\mathcal H}_{g-i}$, its next goal must be a local peripheral node in ${\mathcal P}_{g-(i+1)}$ belonging to the subgraph $G_{g-(i+2)}^{(1)}$ that is one of the three components of $G_{g-i}$, then it continues to jump to the main hub of a subgraph $G_{g-(i+2)}$ that is in the central component of $G_{g-i}$. In this way, the walker moves on until it reaches the trap.

According to the above definitions, it is necessary to introduce two more quantities $p_{g}(i)$ and $h_g(i)$. The former is the MFPT from a node in ${\mathcal P}_{g-i}$ to any of its neighboring nodes in ${\mathcal H}_{g-(i+1)}$ and the latter is the MFPT from a node in ${\mathcal H}_{g-i}$ to any of its neighbors simultaneously belonging to ${\mathcal P}_{g-(i+1)}$ and $G_{g-(i+2)}^{(1)}$. In Appendix~\ref{AppB}, we provide the detailed derivation for $p_g(i)$ and $h_g(i)$, which read
\begin{equation}\label{APPB6}
p_g(i)=2^{i+3}\left[\left(\frac{3}{2}\right)^g-1\right]-\frac{8}{9}\left(\frac{3}{2}\right)^{g-i}+1.
\end{equation}
and
\begin{equation}\label{APPB7}
h_g(i)=2^{i+4}\left[\left(\frac{3}{2}\right)^{g}-1\right]-\frac{16}{9}\left(\frac{3}{2}\right)^{g-i} + 3\,
\end{equation}
respectively.

After obtaining the expressions of related quantities, we next determine the MFPT $T_{h,f}(g)$  from the main hub to the leftmost farthest node, as well as the ATT $T_{f}(g)$.

\subsubsection{Exact solution and leading scaling for the MFPT from the root to the leftmost farthest node}

In order to find the explicit formulae for $T_{h,f}(g)$. We distinguish two cases: (i) $g$ is odd and (ii) $g$ is even.

When $g$ is odd, the target belongs to ${\mathcal P}_1$. In this case, we have
\begin{eqnarray}\label{L5}
T_{h,f}(g)&=&\left[T_{h,p}(g)+\frac{1}{2}T_{\beta_g,{\mathcal B}_{g-1}}(g)\right]\nonumber\\
&\quad&+\sum_{k=0}^{\frac{g-1}{2}-1}p_g(2k)+\sum_{k=0}^{\frac{g-1}{2}-1}h_g{(2k+1)}.
\end{eqnarray}
By plugging Eqs.~(\ref{II11}),~(\ref{III6}),~(\ref{APPB6}), and~(\ref{APPB7}) into Eq.~(\ref{L5}), we obtain a closed-form solution to $T_{h,f}(g)$ given by
\begin{equation}\label{L6}
T_{h,f}(g)=\frac{20}{3}(3^g-2^g)-\frac{176}{15}\left(\frac{3}{2}\right)^g+2g+\frac{179}{15}.
\end{equation}

When $g$ is  even, the target is in ${\mathcal H}_1$. In this case, $T_{h,f}(g)$ can be calculated by
\begin{eqnarray}\label{L7}
T_{h,f}(g)&=&\left[T_{h,p}(g)+\frac{1}{2}T_{\beta_g,{\mathcal B}_{g-1}}(g)\right]\nonumber\\
&\quad&+\sum_{k=0}^{\frac{g}{2}-1}p_{g}(2k)+\sum_{k=0}^{\frac{g}{2}-2}h_{g}(2k+1).
\end{eqnarray}
Substituting Eqs.~(\ref{II11}),~(\ref{III6}),~(\ref{APPB6}), and~(\ref{APPB7}) into Eq.~(\ref{L7}), after some algebra, Eq.~(\ref{L7}) is solved to yield the explicit expression for $T_{h,f}(g)$, given by
\begin{equation}\label{L8}
T_{h,f}(g)=\frac{16}{3}(3^g-2^g)-\frac{176}{15}\left[\left(\frac{3}{2}\right)^g-1\right]+2g.
\end{equation}

Equations~(\ref{L6}) and~(\ref{L8}) can be expressed, respectively, in terms of the network size $N_g$ as
\begin{small}
\begin{eqnarray}\label{L9}
T_{h,f}(g)&=&\frac{20}{3}[N_g-(N_{g})^{\ln2/ \ln3}]+\frac{176}{15}(N_{g})^{1-\ln2/ \ln3}\nonumber\\
&\quad&+\frac{2\ln N_{g}}{\ln3} + \frac{179}{15}
\end{eqnarray}
\end{small}
and
\begin{small}
\begin{eqnarray}\label{L10}
T_{h,f}(g)&=&\frac{16}{3}[N_g-(N_{g})^{\ln2/ \ln3}]+\frac{176}{15}[(N_{g})^{1-\ln2/ \ln3}-1]\nonumber\\
&\quad&+\frac{2\ln N_{g}}{\ln3}\,,
\end{eqnarray}
\end{small}
both of which indicate that for large networks, i.e., $N_g\to\infty$,
\begin{equation}\label{L11}
T_{h,f}(g)\sim N_g\,,
\end{equation}
behaving as a linear function of the network size.

\subsubsection{Closed-form formula and dominating scaling for ATT}

Before evaluating the quantity $T_{f}(g)$, we introduce two new quantities $T_{p,f}(g,n)$ and $T_{h,f}(g,n)$, which denote the MFPT to the leftmost farthest node for a walker starting, separately, from a local peripheral node in ${\mathcal P}_{g-n}$ and a local hub node in ${\mathcal H}_{g-n}$, which is part of the walking path from the root node to the leftmost farthest node. These two intermediary quantities and the interesting quantity $T_{f}(g)$ can be determined by distinguish odd and even $g$.

When $g$ is odd, we have
\begin{equation}\label{L12}
T_{p,f}(g,n) =\sum \limits_{i=\frac{n}{2}}^{\frac{g-1}{2}-1} p_{g}(2i)+\sum\limits_{i=\frac{n}{2}}^{\frac{g-1}{2}-1} h_g(2i+1) \,,
\end{equation}
since for odd $g$, only when $n$ is even, the quantity $T_{p,f}(g,n)$ is meaningful.
Plugging Eqs.~(\ref{APPB6}), and~(\ref{APPB7}) into Eq.~(\ref{L12}) leads to
\begin{eqnarray}\label{L13}
T_{p,f}(g,n) &=& \frac{20}{3}\left(3^{g}-2^{g}\right)-\frac{40}{3}2^{n}\left(\frac{3}{2}\right)^{g} + \frac{56}{15} \left(\frac{3}{2}\right)^{g-n} \nonumber\\
&\quad&+ \frac{40}{3}2^n + 2 g-2 n + \frac{18}{5} \,.
\end{eqnarray}
Analogously, for $T_{h,f}(g,n)$ we have
\begin{equation}\label{L14}
T_{h,f}(g,n) =\sum\limits_{i=\frac{n+1}{2}}^{\frac{g-1}{2}-1} p_g(2i) +\sum\limits_{i=\frac{n-1}{2}}^{\frac{g-1}{2}-1} h_g(2i+1)  \,,
\end{equation}
which, combining Eqs.~(\ref{APPB6}), and~(\ref{APPB7}) into Eq.~(\ref{L12}), gives
\begin{eqnarray}\label{L15}
T_{h,f}(g,n) &=& \frac{20}{3} \left(3^{g} - 2^{g}  \right) - \frac{32}{3} 2^{n} \left(\frac{3}{2}\right)^{g}- \frac{64}{3} \left(\frac{3}{2}\right)^{g-n} \nonumber\\
&\quad&+ \frac{32}{3} 2^{n} + 2 g-2 n + \frac{69}{15}  \,.
\end{eqnarray}
Then, the ATT $T_{f}(g)$ can be determined as
\begin{small}
\begin{eqnarray}\label{L16}
T_{f}(g)&=&  \frac{1}{3} \left[T_{h}(g-1)+T_{h,f}(g)\right]\nonumber\\ &\quad&+\frac{1}{3}\left[T_{p}(g-1)+T_{p,h}(g)+T_{h,f}(g)\right] \nonumber\\
&\quad&+\frac{1}{3^g} \Bigg \{\sum\limits_{i=0}^{\frac{g-1}{2}-1}2\times3^{g-2i-2}\left[T_{p}(g-2i-2)+T_{p,f}(g,2i)\right]\nonumber\\ &\quad&+\sum\limits_{i=0}^{\frac{g-1}{2}-1}
\big[3^{g-2i-3} \left(T_{h}(g-2i-3)+T_{h,f}(g,2i+1)\right)\nonumber\\
&\quad&+3^{g-2i-3}\big(T_{p}(g-2i-3)+T_{p,h}(g-2i-2)\nonumber\\
&\quad&+T_{h,f}(g,2i+1)\big)
\big] + \frac{1}{9} \Bigg\}
 \,.
\end{eqnarray}
\end{small}
The three terms on the rhs of Eq.~(\ref{L12}) account for, respectively, the contribution of nodes in primal $G_{g-1}$, $G_{g-1}^{(1)}$, and $G_{g-1}^{(2)}$ that form $G_{g}$. Using above obtained related quantities, Eq.~(\ref{L12}) is solved to obtain
\begin{small}
\begin{equation}\label{L17}
T_{f}(g)=\frac{20}{3} \left(3^{g} - 2^{g}\right) -\frac{656}{45} \left(\frac{3}{2}\right)^{g} -\frac{28}{9} \left(\frac{2}{3}\right)^{g} +\frac{10}{9}g + \frac{907}{45} \,.
\end{equation}
\end{small}
which can be expressed in terms of network size $N_g$ as
\begin{small}
\begin{eqnarray}\label{L18}
T_{f}(g)&=& \frac{20}{3}\left[N_g-\left(N_g\right)^{\ln 2/\ln 3}\right] -\frac{656}{45} \left(N_g\right)^{1-\ln 2/\ln 3} \nonumber\\
&\quad&-\frac{28}{9} \left(N_g\right)^{-1+\ln 2/\ln 3} +\frac{10}{9}\frac{\ln N_g}{\ln 3} + \frac{907}{45}
 \,.
\end{eqnarray}
\end{small}

Similar to the case of odd $g$, for even $g$ we can obtain
\begin{small}
\begin{equation}\label{L19}
T_{f}(g)=\frac{16}{3}\left(3^{g}-2^{g}\right) - \frac{656}{45} \left(\frac{3}{2}\right)^g-\frac{32}{5} \left(\frac{2}{3}\right)^g+\frac{10}{9}g + \frac{934}{45}
 \,
\end{equation}
\end{small}
and
\begin{small}
\begin{eqnarray}\label{L20}
T_{f}(g)&=& \frac{16}{3}\left[N_g-\left(N_g\right)^{\ln 2/\ln 3}\right] -\frac{656}{45} \left(N_g\right)^{1-\ln 2/\ln 3} \nonumber\\
&\quad&-\frac{32}{5} \left(N_g\right)^{-1+\ln 2/\ln 3} +\frac{10}{9}\frac{\ln N_g}{\ln 3} + \frac{934}{45}
 \,.
\end{eqnarray}
\end{small}
Both Eqs.~(\ref{L18}) and~(\ref{L20}) show that for large networks, i.e., $N_g\to\infty$,
\begin{equation}\label{L21}
T_{f}(g) \sim N_g\,,
\end{equation}
behaving as a linear function of the network size, a scaling identical to that of $T_{h,f}(g)$.
Notice that for odd and even $g$, the degree of the farthest node is 1 and 2, respectively. Thus, $T_{f}(g)$ scales proportionally to the network size and the reciprocal of the degree of the farthest as the trap.


The phenomenon that the leading term of $T_{f}(g)$ displays the same behavior as that of $T_{h,f}(g)$ can be explained based on the following heuristic arguments.
Note that $G_g$ contains three subgraphs, each of which is a copy of $G_{g-1}$. For those nodes in the central subgraph, their MFPT to leftmost farthest node is equal to $T_{h}(g-1)+T_{h,f}(g)$, the dominating term of which is $T_{h,f}(g)$; for nodes in the subgraphs $G_{g-1}^{(2)}$, their MFPT to the leftmost farthest equals  $T_{p}(g-1)+T_{p,h}(g)+T_{h,f}(g)$, the leading term of which is also $T_{h,f}(g)$; while for nodes in the subgraphs $G_{g-1}^{(1)}$ encompassing the leftmost farthest node, their MFPT to the target is less than $T_{h,f}(g)$. Therefore, for all nodes in $G_g$, the dominating term of the ATT $T_{f}(g)$ is proportional to network size $N_g$, which is similar to that of $T_{h,f}(g)$.

\subsection{Result analysis}

In the preceding text we have studied four representative cases of trapping problems with the trap located at the main hub, a peripheral node, a neighbor node of the root with a single degree, and a farthest node from the root, respectively. We have shown that their trapping efficiency exhibits rich behavior.  For the four cases of trapping problems, the leading term of ATT exhibits evidently different dependence on the network size. It can grow sublinearly or linearly with the network size, or behaves as a linear function of network size by a logarithmic correction, which shows that the degree of trap node plays an important role in the trapping efficiency. Our results also demonstrate that for the four cases of trapping problems, the transport processes are very efficient with the ATT increases at most linearly with the network size.

Although for the four trapping problems, the trapping efficiency displays distinct scalings with the network size, Eqs.~(\ref{II233}),~(\ref{III122}),~(\ref{IV6}), and~(\ref{L21}) indicate that the leading scaling of ATT for all four cases grows inversely proportional to the degree of the trap, regardless of its position. For example, for the two cases of trapping problems when the trap is fixed on a peripheral node or a farthest node, the dominating scaling of ATT is identical.
In fact, extensive numerical simulations also verify that for all trapping problems in the hierarchical scale-free network with a single trap, as long as the degree of the trap is identical, the leading behavior for their trapping efficiency is also the same.

It has been proved~\cite{LiJuZh12} that for trapping problem in a general sparse network having $N$ nodes with node $j$ being the trap, the scaling of the lower bound for ATT $T_j$ varies with the network size $N$ as $T_j \sim N/d_j$, where $d_j$ is the degree of the trap node $j$. For the four cases of trapping problems in the hierarchical scale-free network, this minimal scaling for ATT can all be achieved. In most of previously studied networks, this minimum scaling cannot be reached. For example, for trapping in the $(1,3)-$flower and the $(2,2)-$flower with the same degree sequence, when the trap is located on a largest node, their ATT display distinct behaviors, but both are very larger than the minimum scaling~\cite{ZhXiZhLiGu09,ZhYaGa11}. In this sense, the hierarchical scale-free network being studied exhibits the most efficient configuration (the optimal structure) for random walks with a perfect trap fixed at a given node.

\section{Conclusions}

Previous works have shown that for isotropic random walks in general networks in the presence of a single deep trap, the leading scaling for the least ATT is proportional to the size of the network and the inverse degree of the trap. In this paper, we have presented an in-depth analysis on four particular cases of trapping problems in a hierarchical scale-free network, with the trap being located at the main hub, a peripheral node, a neighbor node of the main hub with a single degree, and a farthest node, respectively. For all these four cases, we have derived closed-form formulae for the ATT, as well as their dominating scalings that are all equal to the predicted minimal scalings. In this context, the network under consideration has an optimal structure that is advantageous to efficient trapping.

Our work may have practical implications for designing networks, especially scale-free networks, where minimizing the transport efficiency is a central goal. For example, with the emergency of new preparation techniques, now it is possible to prepare new synthesized polymeric materials with a very complex geometry, e.g., hierarchical scale-free topology, which is favorable to transportation and diffusive dynamics and thus can be utilized as potential artificial antenna systems for light harvesting. Moreover, our technique for computing the ATT is relatively general, which also applies to other networks. For instance, for trapping in dendrimers with a trap fixed on a peripheral node, the ATT can be easily determined by using our method.

\subsection*{Acknowledgment}

This work was supported by the National Natural Science Foundation
of China under Grant Nos. 61074119 and 11275049.

\appendix

\section{Determination of quantities $T_{p,h}(g)$ and $T_{h,p}(g)$} \label{AppA}

According to the particular construction of the hierarchical scale-free network, we can find that the two quantities $T_{p,h}(g)$ and $T_{h,p}(g)$ obey the following recursive relations:
\begin{equation}\label{APPA1}
T_{p,h}(g)=\frac{1}{g}\left[1+\sum_{i=1}^{g-1}\big(1+T_{h,p}(i)+T_{p,h}(g)\big)\right]
\end{equation}
and
\begin{equation}\label{APPA2}
T_{h,p}(g)=\frac{1}{\sum_{i=1}^{g}2^i}\left[2^{g}+\sum_{i=1}^{g-1}2^{i}\big(1+T_{p,h}(i)+T_{h,p}(g)\big)\right].
\end{equation}
The two terms on the right-hand side (rhs) of Eq.~(\ref{APPA1})
can be explained as follows. The first term is based on the fact that the walker takes only one time step to first reach the root. The second term describes the fact that the walker first makes a jump to a local hub node belonging to $\mathbb{H}_i$, then takes
$T_{h,p}(i)$ time steps, starting off from the local hub, to reach any node
in $\mathbb{P}$, and continues to jump $T_{p,h}(g)$ more steps to
reach the root for the first time.
Analogously, the two terms on the rhs of Eq.~(\ref{APPA2}) are based on the following two processes. The first term describes the fact that the walker,
starting from the root, requires only one time step to hit a
peripheral node. The second term explains such a process that the walker,
starting off from the root, first jumps to a local peripheral
node belonging to $\mathbb{P}_i$ in one time step, then makes $T_{p,h}(i)$ jumps to the root, and
proceeds to any node in $\mathbb{P}$, taking $T_{h,p}(g)$ more time steps.

Equations~(\ref{APPA1}) and~(\ref{APPA2}) can be recast as
\begin{equation}\label{APPA3}
T_{p,h}(g)=g+\sum_{i=1}^{g-1}T_{h,p}(i)
\end{equation}
and
\begin{equation}\label{APPA4}
T_{h,p}(g)=2-\frac{1}{2^{g-1}}+\frac{1}{2^{g}}\sum_{i=1}^{g-1}2^{i}T_{p,h}(i),
\end{equation}
respectively. From Eqs.~(\ref{APPA3}) and~(\ref{APPA4}), we obtain
\begin{equation}\label{APPA5}
T_{p,h}(g+1)=g+1+\sum_{i=1}^{g}T_{h,p}(i)
\end{equation}
and
\begin{equation}\label{APPA6}
T_{h,p}(g+1)=2-\frac{1}{2^{g}}+\frac{1}{2^{g+1}}\sum_{i=1}^{g}2^{i}T_{p,h}(i).
\end{equation}
Equation~(\ref{APPA5}) minus Eq.~(\ref{APPA3}) leads to
\begin{equation}\label{APPA7}
T_{p,h}(g+1)-T_{p,h}(g)=1+T_{h,p}(g).
\end{equation}
Similarly, Eq.~(\ref{APPA6}) minus Eq.~(\ref{APPA4}) times $1/2$ yields
\begin{equation}\label{APPA8}
T_{h,p}(g+1)-\frac{1}{2}T_{h,p}(g)=1+\frac{1}{2}T_{p,h}(g).
\end{equation}
Applying the initial conditions $T_{p,h}(1)=1$ and $T_{h,p}(1)=1$, we can solve Eqs.~(\ref{APPA7}) and~(\ref{APPA8}) to arrive at the explicit formulas for $T_{p,h}(g)$ and $T_{h,p}(g)$, which are given by
Eqs.~(\ref{II10}) and~(\ref{II11}), respectively.


\section{Computation of quantities $p_g(i)$ and $h_g(i)$} \label{AppB}

According to the structure of $H_g$, $p_g(i)$ and $h_g(i)$ satisfy the following relations:
\begin{eqnarray}\label{APPB1}
p_g(i)&=&\frac{1}{g-i}\Bigg\{1+[h_{g}(i-1)+p_g(i)+1]\nonumber\\
&\quad&+\sum_{k=1}^{g-i-2}\left[1+T_{h,p}(k)+p_g(i)\right]\Bigg\}
\end{eqnarray}
and
\begin{small}
\begin{eqnarray}\label{APPB2}
&\quad& h_g(i)\nonumber\\
&=&\frac{2}{2(2^{g-i}-1)}\Bigg\{2^{g-i}[1+p_g(i-1)+h_g(i)]+\frac{1}{2}\times 2^{g-i-1}\nonumber\\
&\quad&+\frac{1}{2}\times2^{g-i-1}[1+T_{p,h}(g-i-1)+h_g(i)]\nonumber\\
&\quad&+\sum_{k=1}^{g-i-2}2^k [1+T_{p,h}(k)+h_g(i)] \Bigg\}.
\end{eqnarray}
\end{small}
Equation~(\ref{APPB1}) can be explained as follows. The first term describes the process that the walker starting off from a node in ${\mathcal P}_{g-i}$ may directly go to the target. Alternatively, the walker can jump to the neighboring node in ${\mathcal H}_{g-(i-1)}$, then takes time $h_{g}(i-1)$ to reach one of the local peripheral nodes in ${\mathcal P}_{g-i}$, and continues to bounce $p_g(i)$ steps to hit the target; this process is explained by the second term. The last term represents the fact that the particle goes to a local hub in ${\mathbb H}_{c}$ $(1\leq c\leq g-i-2)$, from which it takes an average steps $T_{h,p}(k)$ to return to one of the local peripheral nodes in ${\mathcal P}_{g-i}$, and then moves on average $p_g(i)$ steps to arrive at the target. Analogously, we can explain Eq.~(\ref{APPB2}).

Merging similar terms of Eqs.~(\ref{APPB1}) and~\ref{APPB2}) leads to
\begin{equation}\label{APPB3}
p_g(i)=h_g(i-1)+g-i+\sum_{k=1}^{g-i-2}T_{h,p}(k)
\end{equation}
and
\begin{eqnarray}\label{APPB4}
h_g(i)&=&4p_g(i-1)+\frac{1}{2^{g-i-2}}\sum_{k=1}^{g-i}2^k+T_{p,h}(g-i-1)\nonumber\\
&\quad&+\frac{1}{2^{g-i-2}}\sum_{k=1}^{g-i-2}2^k T_{p,h}(k).
\end{eqnarray}
Substituting Eq.~(\ref{APPB4}) into Eq.~(\ref{APPB3}), we can obtain the recursive relation for $p_g(i)$ as
\begin{eqnarray}\label{APPB5}
p_g(i)&=&4p_g(i-2)+\frac{1}{2^{g-i-1}}\sum_{k=1}^{g-i+1}2^k+T_{p,h}(g-i)+\nonumber\\
&\quad&+\frac{1}{2^{g-i-1}}\sum_{k=1}^{g-i-1}2^{k} T_{p,h}(k)+(g-i)+\sum_{k=1}^{g-i-2}T_{h,p}(k).\nonumber\\
\end{eqnarray}
Considering the initial conditions $p_g(0)=64/9\times(3/2)^g-7$ and applying Eqs.~(\ref{II10}) and~(\ref{II11}), Eq.~(\ref{APPB5}) can be solved to yield
\begin{equation}\label{APPB6B}
p_g(i)=2^{i+3}\left[\left(\frac{3}{2}\right)^g-1\right]-\frac{8}{9}\left(\frac{3}{2}\right)^{g-i}+1.
\end{equation}
Plugging Eq.~(\ref{APPB6B}) into Eq.~(\ref{APPB4}), we obtain the analytical expression of $h_g(i)$ as
\begin{equation}\label{APPB7B}
h_g(i)=2^{i+4}\left[\left(\frac{3}{2}\right)^{g}-1\right]-\frac{16}{9}\left(\frac{3}{2}\right)^{g-i} + 3.
\end{equation}

\end{document}